Pérez Ramírez, Rigoberto
Privatizaciones, fusiones y adquisiciones: las grandes empresas en México
Espacios Públicos, vol. 16, núm. 37, mayo-agosto, 2013, pp. 113-140
Universidad Autónoma del Estado de México
Toluca, México

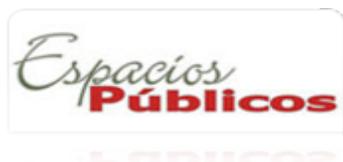

# Privatizaciones, fusiones y adquisiciones: las grandes empresas en México

Privatizations, mergers and acquisitions: large companies

*Rigoberto Pérez Ramírez*\*

**Resumen**

*Se analiza la evolución del proceso de privatización, fusiones y adquisiciones de las grandes empresas en el país en las últimas décadas, para comprender los hilos conductores que configuraron los cambios estructurales de la economía, a fin de insertarla al mercado global caracterizado por formaciones de alianzas estratégicas oligopólicas mundiales, a través de las fusiones y adquisiciones que privilegian a las empresas transnacionales.*

Palabras clave: privatización, fusión, adquisición.

**Abstract**

The present work has as principal objective analyze the evolution of the process of privatization, mergers and acquisitions of the big companies in the country in the last decades, to understand the conductive threads that formed the structural changes of the economy, in order world oligopólicas to insert it to the global market characterized by formations of strategic alliances, across the mergers and acquisitions that they favour to the transnational companies.

\* Universidad Autónoma del Estado de México, México / rferezram@yahoo.com.mx

## EL CONTEXTO

En las últimas décadas la economía mundial se ha visto transformada de manera más rápida e intensa, evidenciando cambios estructurales en la organización y funcionamiento del propio capitalismo moderno. Este nuevo capitalismo muestra los impactos de las tendencias que se expresan en el proceso de la globalización, tanto en los ámbitos de la economía, como en la política y la cultura.

El proceso de globalización de la economía mundial, se refiere a la tendencia de liberalización de los mercados de los flujos internacionales de capital, dinero y mercancías; la internacionalización de la producción bajo el monopolio de las empresas transnacionales favorecidas por la privatización; la revolución tecnológica vinculada a las modernas posibilidades de elaboración y transferencias de información; el comercio global y el manejo de los mercados por formaciones de alianzas estratégicas oligopólicas mundiales, a través de las fusiones y adquisiciones.

Esta nueva configuración económica del mapa mundial agudiza la competencia internacional de las economías nacionales, que ajustan sus políticas a la dinámica del mercado global[1] y a las estrategias de las empresas globales transnacionales. Es decir, se impone un paradigma que se sustenta en la economía de mercado con un marcado protagonismo de las transnacionales favorecidas por las privatizaciones, cuestionando las concepciones de la teoría y de la política económica del Estado nación[2], y por ende, debilitando a la empresa pública y al Estado interventor.

Cuando se habla de la modificación estructural del capitalismo moderno que se origina con la crisis económica mundial en la década de los setenta, la cual persiste hasta nuestros días, se comprende con mayor precisión el significado de globalización. En otras palabras, la quiebra del sistema monetario internacional *Bretton Woods* –bajo el dominio del dólar estadounidense como dinero regulador de la economía internacional–, abre el camino a la revalorización del capital en la esfera financiera y acelera el proceso de concentración y centralización de capitales que la propia crisis alimenta; es decir, es la liberalización de los mercados de los flujos internacionales de capital, dinero y mercancías; como condición previa de la renovada implementación tecnológica en el proceso de producción capitalista, para la apertura de nuevos mercados de carácter global (Hirsh, 1996).

De ahí, que la globalización sea en esencia un proceso económico, pero cuya complejidad no se reduce a la inevitable lógica del capital, sino como señala Hirsch, es además una estrategia política pues el capital se internacionaliza, "en coordinación con los gobiernos neoliberales que, a consecuencia de la crisis, llegaron al poder. La política económica de liberalización, privatización y desregulación tiene como meta crear las condiciones políticas institucionales adecuadas para una transformación en la correlación de fuerzas de las clases, tanto nacional como internacional; he aquí la condición para la reorganización técnica de la producción capitalista" (Hirsh, 1996: 90).

En la últimas décadas de finales del siglo XX y principios del XXI, las políticas de liberalización financiera[3] han provocado serios problemas de falta de liquidez e insolvencia en el ámbito monetario y financiero de las distintas economías nacionales debido al proceso denominado globalización financiera,[4] que se caracteriza por el libre flujo de capitales, una concentración de éstos en grupos oligopólicos, elevada especulación y excesivos márgenes de intermediación financiera externa en los mercados nacionales, al margen de los canales y reglas del juego de la institucionalidad monetaria y financiera internacional.

La inestabilidad del sistema monetario y financiero internacional se evidenció más nítidamente con la crisis de los mercados emergentes.[5] Primero, la crisis mexicana en 1994, que amenazó con propagarse a las economías de Argentina y Brasil, denominada por ello, *Efecto Tequila*. Luego, la crisis asiática en 1997, llamada *Efecto Dragón*, que abarcó los países de Indonesia, Malasia, Filipinas, Tailandia y Corea del Sur y, se prolongó a Hong Kong (incorporado en 1999 a la República Popular China) y a la crisis bancaria de Japón, hasta 1998.

En la primera década del 2000, la crisis estuvo caracterizada por burbujas especulativas financieras que afectaron a las empresas *Enron Corp.* (energéticos) y *WorldCom Inc.* (comunicaciones), en el 2001 y 2002, respectivamente, cuestionando el modelo de regulación financiera estadounidense. Sin embargo, en 2007, la crisis financiera se agudizó con el colapso del mercado inmobiliario de hipotecas de alto riesgo "*subprime*" (préstamos para vivienda otorgados a las familias de bajos ingresos que no cumplían con los requisitos para obtener hipotecas) de los Estados Unidos, cobrando intensidad con la quiebra de *Lehman Brothers* en septiembre de 2008, y expandiéndose a los bancos de la zona euro que habían asegurado muchos de los títulos de créditos en ese país. Según estimaciones del Fondo Monetario Internacional (FMI), "las pérdidas totales registradas por los bancos de la zona euro entre 2007 y 2010, asciende a 630 000 millones de dólares estadounidenses, no lejos de los 878 000 millones de los bancos americanos" (FMI, 2010:12).

La crisis que se extendió a todas las economías del mundo en 2008, también golpeó a América Latina. La media ponderada del Producto Interno Bruto (PIB) se redujo cerca del 1.9 por ciento, mientras que el PIB por habitante cayó casi 3 %. Uno de los países más perjudicados fue México, donde el PIB se contrajo cerca de 6.5% en 2009, y el PIB por habitante cayó un 7.5%. Mientras que los salarios medios mensuales cayeron de 2.0 a 5.0 % durante 2008-2009 (OIT, 2010)

En México, a raíz del agotamiento del modelo intervencionista estatal y la crisis de la deuda externa en 1982, los distintos gobiernos han desmantelado gradualmente el modelo y siguen manteniendo una continuidad en la política económica de privatización, ya sea como estrategia de cambio estructural, modernización o reforma estructural.

En la década de los ochenta, el cambio estructural de la economía como eje ordenador de su modernización, comprendía la reordenación del sector público, su reconversión industrial y la estimulación del sector privado como pivote de la producción y saneamiento de las finanzas públicas, con el fin de dinamizar e insertar nuestra economía al mercado global.

La privatización en México continua siendo un asunto relevante en la agenda nacional, particularmente en momentos en los que el retorno del PRI a la presidencia de la república pondrá en la mesa del debate del Congreso de la Unión las reformas neoliberales pendientes (energética, laboral y fiscal) que necesita el país para transitar a un estadío de mejor desarrollo económico, político y social, y hacer frente con mayor capacidad de respuesta a los distintos escenarios de la globalización, que en la actualidad se caracteriza por las megafusiones y adquisiciones de las empresas. En este contexto, para el presente artículo se retomarán las aportaciones de algunos autores como José Ayala Espino, Jacques Rogozinski, Kenneth Davidson, José Luis Calva, Víctor Livio de los Ríos, John Saxe-Fernández, Celso Garrido, que han revisado a la literatura y estudios del tema que se aborda.

## CONCEPTUALIZACIÓN DE PRIVATIZACIÓN

La privatización, más que la trasferencia de las empresas públicas del Estado a manos del sector privado, es una estrategia de la globalización de los mercados, en la que convergen el avance de la concentración económica y el reparto de mercados considerando una economía mundial con bloques y regiones.

La privatización predomina en el paradigma de la economía de mercado, que parte del supuesto de que el mercado es eficiente *per se*, por lo que deben ser transferidas a las fuerzas del mercado las funciones del Estado.

La mayoría de los autores coinciden en que el término "privatización" es complejo, dada la diversidad de significados que lo hacen un concepto no-unívoco. Son variadas las formas en que el fenómeno es interpretado: técnico, económico, social, político y jurídico, pero si nos ajustamos a la definición del diccionario inglés, la palabra privatizar (*privatize*), que aparece por primera vez en 1983, significa "la transferencia de las instalaciones, infraestructura y servicios del Estado al sector privado" (Hanke, 1989: 81-89).

La privatización supone "la existencia de un Estado mínimo y la integración de los principios de libre competencia, disciplina de mercado y realidad económica en la mayoría de las funciones públicas e incluso de algunas consideradas tradicionalmente como expresiones de la soberanía estatal, tales como el sistema penitenciario o el orden público" (Almada, 1989: 18-19).

Para Naranjo, la privatización "se inscribe dentro de [*lo que se ha*] denominado 'desestatización' que alude al proceso de desplazamiento de las actividades del sector público hacia el sector privado, mediante la sustitución de la gestión estatal tradicional de organismos públicos por una gestión basada en criterios de mercado"[6] (Naranjo, 1992: 12).

Para Ayala Espino (1997), la privatización es un concepto tanto económico como político, y tiene por supuesto implicaciones ideológicas diversas.

La privatización busca desde la economía hacer eficiente al Estado transfiriendo las empresas públicas al sector privado a través de la venta, fusión, extinción y liquidación, para garantizar la producción, la inversión y la competitividad.

Bajo la óptica política, la privatización busca contrarrestar y/o revertir la centralización en la toma de decisiones, el patrimonialismo, el rentismo y las ineficiencias en la prestación de servicios públicos. Para Rodríguez (1991), en dicha perspectiva existen dos objetivos principales. El primero, se ocupa de redefinir las esferas pública y privada que permita una participación amplia de los ciudadanos en la toma de decisiones del capital, una "democratización del capital", y en general una reforma democrática del sistema político. El segundo propone establecer nuevas condiciones de competencia política haciendo a un lado las viejas coaliciones que apoyan las políticas intervencionistas por medio de presiones corporativas o rentistas.

Desde el punto de vista ideológico, la privatización significa que el Estado es demasiado obeso con una administración pública burocrática e ineficiente, por lo que debe redimensionarse para lograr la eficiencia que establece la disciplina del mercado.

Estas acepciones de privatización que desplazan el papel protagonista del Estado en la economía, reduciendo el tamaño de su sector público,[7] no resuelven los problemas de fondo. Por lo tanto, la interpretación neoliberal de un Estado moderno reduciendo el tamaño del mismo, no resuelve los problemas de raíz. En lugar de ver en el tamaño, el efecto y no la causa, debe redefinirse su función o, bien, reinterpretar su papel a la luz de nuestro tiempo.

**FORMAS DE PRIVATIZACIÓN**

Para diversos autores, la venta de las empresas no es la única forma posible de privatizar la producción pública. Existen algunas alternativas que, en ciertos casos, pueden tener mayores posibilidades de implementación que la venta directa.

Horacio Boneo (1985), identifica tres alternativas:
a) La "privatización de gerencia", en el sentido de lograr un adecuado balance entre el componente "público" y el "empresarial" de una firma estatal, que en muchos casos parece estar sesgado hacia el primer componente.

b) "Privatización periférica", es decir, la venta de ciertas partes de la empresa, manteniendo el núcleo productivo bajo la propiedad estatal.
c) La privatización por medio de acuerdos y convenios con socios provenientes del sector privado, tanto nacionales como transnacionales.

Para Feigenbaum y Hening (1994), la privatización no se reduce a la simple venta o transferencia de empresas públicas al sector privado, sino que es abarca más alcanzando aspectos más sutiles, menos evidentes, a la cual denominan "privatización sistemática", que reconfigura la sociedad en su conjunto, alterando las instituciones y los intereses económicos y políticos. Esta forma de privatización trata de disminuir las expectativas de la sociedad con relación a la responsabilidad del Estado, reducir el mantenimiento y apoyo de la infraestructura por parte del sector público, y transformar el mosaico de grupos de interés para hacerlo menos proclive a apoyar el crecimiento del aparato del Estado.

Rodríguez Arana (1991) conceptualiza la privatización como una devolución de tareas o servicios realizados por empresas públicas a empresarios privados. Considera que en la privatización de servicios pueden encontrarse formas muy variadas, entre ellas, la privatización formal y la privatización material.

La privatización formal está compuesta por la privatización económico-financiera y la privatización legal. La económico-financiera es el traspaso al sector privado de la financiación de una tarea pública, por ejemplo, capital privado en proyectos de inversión pública o para ámbitos de servicios, como el caso de construcción de escuelas. En la privatización legal, el Estado sigue siendo titular de la entidad y debe asumir los déficits.

La privatización material se divide en privatización organizativa y privatización funcional. La organizativa asume la responsabilidad del cumplimiento de la tarea, pero la realiza la empresa privada. En esta categoría a su vez caben dos variantes: la forma de submisión en la que el Estado sigue siendo el responsable del servicio, pero recomienda la tarea al sector privado como empresa subsidiaria; y el sistema de concesión que se caracteriza por el principio del riesgo y ventura que asume el contratista, se entrega al sector privado la realización de la tarea y la responsabilidad financiera. En la privatización funcional, la entidad pública también traspasa al sector privado la competencia de la tarea y sólo es responsable subsidiariamente.

Coincido con los autores en la necesidad de construir alternativas que no impliquen como única solución la venta directa de las empresas estatales. Pero, se requiere de la adecuación de las leyes, reglamentos y procedimientos, es decir, de un marco regulatorio apropiado para fomentar la iniciativa pública y privada, y evitar por ejemplo, un monopolio bilateral entre una empresa pública y un proveedor privado; corrupción, o que la empresa pública sea absorbida con el tiempo por la empresa privada reafirmando la privatización. Y con ello, la retracción del Estado en la actividad económica como promotor del desarrollo.

La privatización es el *factotum* que los neoliberales consideran necesario para que cualquier país del orbe pueda transitar a un estadío de mejor desarrollo económico, político y social, y hacer frente con mayor capacidad de respuesta a los distintos escenarios de la globalización, la cual se caracteriza en la actualidad por las megafusiones y adquisiciones de las empresas.

**EL ORIGEN DE LAS FUSIONES Y ADQUISICIONES**

Las fusiones y adquisiciones (F y A, o *M y A, Merger and Acquisition*) son antiguas como la existencia de las empresas, que a lo largo de la segunda mitad del siglo XX alcanzaron mayor relevancia. A partir de la década de los noventa, las operaciones societarias de fusión y adquisición en los distintos sectores económicos, han sido uno de los fenómenos destacados de los mercados financieros internacionales.

Kenneth Davidson (1985) distingue cuatro oleadas de fusiones y adquisiciones que abarcan desde finales del siglo XIX y principios del XXI, a las que se les atribuyen diferentes fuerzas impulsoras.

La primera oleada corresponde a la llamada "integración horizontal" que se extendió entre finales del siglo XIX y principios del XX, cuyo objetivo principal de las empresas era aumentar su capacidad de fabricación, beneficiándose con la generación de economías de escala.

La segunda oleada abarca aproximadamente la década de los veinte del siglo XX y recibió el nombre de "integración vertical", su objetivo era reducir los costos operativos para mantener los márgenes de beneficios, a través del control sobre toda la cadena productiva. Esta oleada terminó con la caída de la bolsa en 1929.

La tercera oleada se conoció como "la era del conglomerado" e inició después de la Segunda Guerra Mundial hasta 1970. Las empresas pretendían liberarse de la reglamentación anti-*trust* y estabilizar su rendimiento financiero, compra empresas con capacidad para contraer mayores deudas y proporcionar suficiente efectivo para llevar adelante nuevas adquisiciones. Sin embargo, este periodo concluyó cuando los inversionistas comenzaron a dividir los conglomerados por la dificultad percibida para la gerencia de controlar efectivamente sus diversos corporativos.

La cuarta oleada inicia en la década de los ochenta y se caracteriza por la transformación industrial, motivando a las empresas a reaccionar ante los rápidos cambios tecnológicos, buscando acceso a nuevas tecnologías que permitan reducir costos, mejorar la productividad y desarrollar nuevos productos. Esta oleada también atestiguó los apoderamientos hostiles.

El apoderamiento se enfocaba a grandes conglomerados que eran desmantelados y vendidos en fragmentos. Estas adquisiciones hostiles eran financiadas normalmente por compras apalancadas,[8] sin embargo, el periodo terminó después de varias quiebras muy publicitadas a causa de las compras apalancadas.

A principios de la década de los noventa y hasta la fecha, las fusiones y adquisiciones se caracterizan por las megafusiones en las que se combinan varios factores como la revolución de la información tecnológica, la desregulación, menores barreras al comercio y la globalización.

Las formas de asociación o alianzas estratégicas entre empresas de diferentes nacionalidades, giros y actividades productivas, representan grandes movimientos de capital así como inversiones de gran magnitud, determinantes en las relaciones económicas internacionales por el enorme poderío de sus relaciones de inversión y expansión.

Los principales sectores económicos con mayor dinamismo son el bancario y financiero, el de telecomunicaciones, servicios, energía eléctrica, gas, distribución de agua, alimentos y sus derivados, farmacéuticos y la minería.

En estos sectores las megafusiones más importantes entre corporativos internacionales en la década de los noventa son: Chase Manhattan y Chemical Bank, Nations Bank y Bank of America, Banco Mitsubishi y Bank of Tokio, Citicorp y Travelers, Deutsche Bank y Dresdner Bank; Exxon y Mobil, British Petroleum y Amoco; AT&T y Tele-Comm; Daimler Benz y Crysler y, Glaxco Wellcome con Smith-K*line Beecham*, entre muchas otras.

Las fusiones y adquisiciones tienen un carácter cíclico por el comportamiento que muestra el proceso general de la economía, por lo que se reactivan una y otra vez. En 2006, *Thomson Financial Mergers* informó que ese año estuvo marcado por una cifra récord en el ámbito de las fusiones y adquisiciones a nivel mundial, alcanzándose 2,6 billones de euros. Esto significó un aumento de 35% respecto a 2005.

Desde la teoría económica y financiera, la lógica de un proceso de fusión y adquisición debe perseguir la creación de riqueza para el accionista, a través de la maximización del valor de la empresa. Este hecho sucederá cuando exista un efecto sinérgico positivo, es decir, cuando el valor de mercado de la empresa resultante, sea mayor que el valor de las dos sociedades independientemente consideradas.

## CONCEPTUALIZACIÓN DE FUSIONES Y ADQUISICIONES

Se llama fusión a la unión de dos o más empresas independientes para formar una nueva entidad superior, cuyo objetivo principal sea incrementar su presencia y dominar en el mercado. Las fusiones pueden ser horizontales o verticales.

Cuando se fusionan dos empresas que venden el mismo bien o el mismo tipo de bien dentro del mercado, se denomina fusión horizontal o lateral. Por ejemplo, tres instituciones de crédito deciden fusionarse y constituir una nueva institución de crédito, donde prevalece el mismo giro. Es usual impedir este tipo de fusiones porque se considera que va contra la libre competencia y tienden a crear mercados oligopólicos y en lo posible monopolios.

Las fusiones de tipo vertical se dan entre empresas que realizan distintas etapas del proceso de producción. Para Taylor (2003), la fusión vertical consiste en integrar empresas que permiten hacer integración vertical hacia adelante (acercamiento a los clientes) o hacia atrás (acercamiento a las materias primas), para integrar en una sola empresa, la cadena productiva de un bien o servicio.

A estas fusiones se suman las diversificadas o conglomeradas y las concéntricas. Las primeras constituyen un conglomerado de empresas, es decir, se adquieren empresas que fabrican productos diferentes y actúan en mercados distintos (Nazlhe, Labatut y Aparisi, 2009). Por su parte, Lorig (1991), observa que las fusiones concéntricas se producen entre firmas que de alguna manera están relacionadas entre sí.

La adquisición se refiere a aquella empresa que compra o adquiere otra, o la división de una empresa, sin que ésta pierda sus características fundamentales. Es una negociación directa, en la que una empresa compra los activos o acciones de la otra y en la que los accionistas de la compañía adquirida, dejan de ser los propietarios de la misma.

Las adquisiciones o *take-overd* representan la compra de una empresa por otra, generalmente de mayor tamaño, pagando un valor superior por el de todas sus acciones juntas dentro del mercado accionario; sin embargo, "… las grandes adquisiciones se llevan a cabo mediante un intercambio accionario o *stocks* de préstamo, que puede incluir el pago de alguna diferencia en efectivo, emitidos por la empresa 'adquiriente'. Una adquisición 'apalancada' o *LBO* (por sus siglas en inglés de *leveraged buy out*) representa una adquisición que ha sido financiada por una gran proporción de crédito de un agente financiero" (García, 2000: 5).

Otra figura de adquisición es la inversa o reverse *take-over*, en la que una empresa de menor tamaño adquiere una de mayor magnitud, algunos de estos casos se presentan mediante la privatización de empresas públicas.

Una vez establecido el marco conceptual de la privatización, la fusión y adquisición de empresas, es necesario estudiar *grosso modo* algunas experiencias de los procesos de ajuste estructural de la economía de nuestro país a la dinámica del mercado global, principalmente, por medio de la venta de empresas públicas al sector privado y, la posterior fusión y adquisición de grandes firmas en México.

**EL PROCESO DE PRIVATIZACIONES, FUSIONES Y ADQUISICIONES DE GRANDES EMPRESAS EN MÉXICO**

A falta de un crecimiento sostenido a mediados de la década de los setenta, que se agudiza en 1982 con la crisis de la deuda externa, el Estado mexicano se ve obligado a desmantelar las entidades paraestatales bajo el supuesto de crecer con bases firmes y permanentes en las décadas posteriores, en la inteligencia de aplicar sin reserva alguna el paradigma neoliberal: liberalización, desregulación y privatización a ultranza.

En este sentido, las empresas transnacionales acordes con su capacidad gerencial, tecnológica, económica y grado de eficiencia expanden sus intereses más allá de las fronteras del propio Estado que las vio nacer, en busca de nuevos mercados para disminuir costos y elevar ganancias, estableciendo alianzas, fusiones, adquisiciones, coinversiones, integraciones y demás formas de asociación, para una mayor participación y control de los mercados, que derivan en la monopolización y oligopolización del sector donde operan. De esta manera, las empresas transnacionales se convierten en el centro vital del proceso moderno de acumulación capitalista, así como de la moderna racionalidad tecno-empresarial que somete las demandas y solicitudes de los particulares a la intervención estatal y desliza la toma de decisiones, ejecución, control y evaluación al *management* de la iniciativa privada.

La privatización y la posterior fusión y adquisición de grandes empresas inician durante el gobierno de Miguel de la Madrid, y se inscriben en cinco modalidades de acuerdo con las características de la empresa, para fortalecer el Estado sólo en aquellas áreas irrenunciables por ser estratégicas y prioritarias, como lo muestra el cuadro 1.

*Cuadro 1*
Mecanismos de desincorporación de entidades paraestatales

| *Características de la empresa* | *Mecanismos* |
|---|---|
| Incumplimiento de los objetivos para los que fue creada<br>No rentable y sin potencial económico<br>Existencia sólo en el papel | Extinción o liquidación |
| Prioritaria y vinculada con los programas de desarrollo regional | Transferencia a gobiernos estatales |
| Incremento en eficiencia al unirse a dos o más entidades paraestatales | Fusión |
| Entidad no estratégica ni prioritaria | Venta o enajenación |

Fuente: Rogozinski (1993: 110).

Rigoberto Pérez Ramírez

Así tenemos que de las 1 155 empresas existentes formalmente en 1982 (103 eran organismos descentralizados,[9] 754 empresas de participación estatal mayoritaria,[10] 75 de participación estatal minoritaria y 223 fideicomisos públicos),[11] de las cuales al 31 de agosto de 1988, se habían sujetado 722 a desincorporación (260 de liquidación, 136 de extinción, 28 de transferencia y 218 de venta) que sumados a las 48 desincorporaciones mediante la entrada en vigor en mayo de 1986 de la Ley de Entidades Paratestatales, arrojan un total de 770. La desincorporación brindó una experiencia invaluable para la privatización de grandes y complejas entidades paraestatales que comenzó en 1989, previo saneamiento financiero de las empresas antes de venderlas, en la administración de Salinas de Gortari.

La privatización junto con otros instrumentos de política económica como la desregulación (eliminación de la aplicación discrecional de las normas que regulan la actividad económica, así como eliminación de la excesiva reglamentación de las barreras de entrada y salida de mercados, y de prácticas monopólicas y oligopólicas), la apertura comercial orientada hacia la exportación y competitividad de la economía, y la inversión extranjera directa como complemento a la inversión nacional, buscó en tercera instancia contribuir a los objetivos de incrementar la eficiencia en los factores de producción e incentivar mayor inversión y crecimiento económico, así como un nivel de bienestar más alto de la población (Rogozinski, 1997).

El número de entidades paraestatales disminuyó de 1 155 en 1982 a 210 en 1993, pero para junio de 1994 hubo un incremento de entidades debido a la creación de las Administraciones Portuarias Integrales (API), con el fin de modernizar el sistema portuario, así el año concluyó con 219 entidades: 82 organismos descentralizados; 107 empresas de participación estatal mayoritaria; y 30 fideicomisos públicos, menos de la mitad de las que se tenían al inicio de la administración. Entre estas empresas destacan Aeroméxico, Mexicana de Aviación, Compañía Minera de Cananea, Sidermex, Teléfonos de México.[12] También se vendieron los 18 bancos, producto de la nacionalización de 1988, que representó un monto global de 36 billones de pesos (Salinas, 1990).

Debido a lo anterior, el sector paraestatal redujo su importancia en términos de empleo y producción totales. La participación de las empresas públicas en la producción estatal se redujo de 25 % en 1983 a menos de 16 % en 1992. El empleo, por su parte, disminuyó su participación casi 10 % respecto de la ocupación total en la economía (Aspe, 1993), como muestra el cuadro 2.

De esta forma, el gobierno de Salinas de Gortari se retiró de 21 ramas de la actividad económica, entre las que destacan: minería, construcción, comercio, transporte, servicios financieros, industria manufacturera y servicios comunales, de las 63 ramas en las que participaban las 1 155 entidades paraestatales.

*Cuadro 2*
Empresas paraestatales vendidas con el mayor número de empleados

| Compañía | Número de Empleados |
|---|---|
| Teléfonos de México | 51 126 |
| Bancomer | 37 041 |
| Banamex | 31 385 |
| Mexicana de Aviación | 13 027 |
| Impulsora de la Cuenca del Papaloapan | 3 617 |
| Astilleros Unidos de Veracruz | 2 988 |
| Compañía Minera de Cananea | 2 973 |
| Compañía Real del Monte y Pachuca | 2 416 |
| Dina Camiones | 1 678 |
| Tabamex | 1 259 |

Fuente: Aspe (1993: 184).

El sector manufacturero en el periodo 1982-1988 y 1989-1994, fue el que más sufrió privatizaciones. Siguieron el mismo camino el sector financiero; transporte, almacenamiento y comunicaciones; minería; electricidad, gas y agua.

La participación de capital extranjero se reduce a unas cuantas firmas, sin duda algunas de ellas importantes, sin que sea el hecho más destacado (Rogozinski, 1993).[13] Entre las empresas en que la participación extranjera es del 100%, y en las que las firmas extranjeras eran anteriormente accionistas, se encuentran los casos de Fermentaciones Mexicanas y la Constructora Nacional de Carros de Ferrocarril. El dato de la tecnología era argumento para realizar esta forma de privatización. En otras compañías, como en Peña Colorada adquieren la participación que tenía el Estado y son socios con otros importantes grupos del país. Incluso en las dos mayores empresas en que participan no tienen el control de las firmas, son los casos de Mexicana de Aviación y de Teléfonos de México. Destaca del grupo tan sólo la compra de la Siderúrgica del Balsas, que es una de las razones sociales en que se dividió la antigua Siderúrgica Lázaro Cárdenas "Las Truchas" (SICARTSA).

El proceso de privatización continuó en el gobierno de Zedillo Ponce de León (1994-2000). En este periodo, la desincorporación del sector paraestatal se dio bajo la forma de concesiones y licitaciones para operar bienes y servicios del sector público, así como la venta de activos (Ortega, 1999).

En el primer semestre de 1995, el presidente Zedillo modificó la Constitución Política, para permitir la participación privada en comunicaciones por satélite, que antes estaba reservada para el Estado, y se aprobó una nueva Ley de Telecomunicaciones. En el mismo sentido se encontró ferrocarriles, petroquímica, gas natural, aeropuertos y puertos, bajo la misma tesis de privatización: con la venta de paraestatales se elimina una de las principales fuentes del déficit fiscal; se reduce la aceleración en el crecimiento de precios; disminuyen las presiones sobre las tasas de interés, y se fomenta la repatriación de capitales, la inversión extranjera directa y una mejor estructura de la deuda pública (Flores, 1997).

Del 1 de diciembre de 1994 al 31 de julio de 2000, concluyó la desincorporación de 111 entidades mediante los siguientes procesos: 31 a través de disolución, liquidación y extinción; 40 mediante la enajenación de la participación accionaria que poseía el gobierno federal o alguna otra entidad paraestatal; tres vía transferencia a gobiernos estatales; 21 de conformidad a la LFEP que dejaron de considerarse como tales; y 16 mediante su fusión.

Entre el 1 de septiembre de 1999 y el 31 de julio de 2000, concluyeron 32 procesos de desincorporación de entidades paraestatales, 15 mediante operaciones de venta de la participación accionaria que poseía el gobierno federal: Satélites Mexicanos; Grupo Aeroportuario del Pacífico; Servicios a la Infraestructura Aeroportuaria del Pacífico; y los Aeropuertos de Aguascalientes, El Bajío, Hermosillo, Guadalajara, Manzanillo, Mexicali, Los Mochis, Morelia, La Paz, Puerto Vallarta, San José del Cabo, y Tijuana, todas S.A. de C.V.; 16 a través del proceso de fusión: distribuidora CONASUPO de Campeche, Hidalgo, Michoacán, Oaxaca, Tamaulipas, Veracruz, El Bajío, Centro, Noroeste, Norte, Pacífico, Sur, Sureste, Metropolitana, Norte-Centro y Peninsular, todas S.A. de C.V.; y una mediante el proceso de disolución y liquidación de Servicios Portuarios de Manzanillo, S.A. de C.V. (Zedillo, 2000).

La racionalidad del Estado a través de la privatización permanece en el gobierno de Vicente Fox, destacando las reformas estructurales de "segunda generación", que tienen como propósito terminar con todo vestigio de la empresa pública (privatización del sector energético), agilizar una reforma laboral que termine de crear un ámbito 100% flexible de contratación y una reforma fiscal.

Al final del gobierno de Vicente Fox, del 1 de enero al 31 de julio de 2006, se crearon dos organismos, el Fondo Nacional para el Consumo de los Trabajadores y la Comisión Nacional de Vivienda, con lo que sumaron 21 entidades públicas creadas a partir del 1° de diciembre de 2000. Deja el gobierno con 215 entidades paraestatales, al 31 de julio de 2006. De ese total, 173 eran organismos, empresas y fideicomisos públicos vigentes, 90 organismos públicos descentralizados, 65 empresas de participación estatal mayoritaria y 18 eran fideicomisos públicos.

Por otra parte, a pesar de las serias dudas del triunfo de Felipe Calderón en las elecciones presidenciales de 2006, éste toma posesión del gobierno con un alto déficit de legitimidad. Hecho que marca todo su sexenio, sin embargo, continúa la política económica de privatización. Del 1° de agosto de 2011 al 23 de julio de 2012, se crearon dos entidades nuevas y no se concluyeron ni iniciaron procesos de desincorporación. Al 23 de julio de 2012, las entidades paratestatales suman 197, de las cuales 185 son organismos descentralizados, empresas de participación estatal mayoritaria y fideicomisos públicos que se encuentran vigentes y 12 en proceso de desincorporación. De las entidades vigentes 99 son organismos descentralizados, 67 empresas de participación estatal mayoritaria y 19 fideicomisos públicos (SHCP, 2012). Estas entidades vigentes representan el 16 % del redimensionamiento de las 1 155 empresas públicas que se tenían en 1982. (Véase anexo cuadro 3).

## LA PRIVATIZACIÓN PERIFÉRICA DE PEMEX

En materia de petróleo, gas y electricidad, la agenda de Fox profundiza los dispositivos puestos en marcha desde los ochenta para desarticulación administrativa y financiera y la inducción de la privatización y extranjerización del complejo petroeléctrico. Por ejemplo, el desfinanciamiento crónico de Pemex-CFE, por la vía de un focalizado ataque fiscal, promueve su creciente endeudamiento. Hoy, al esfuerzo acumulado en tres décadas, Pemex, con una integración debilitada, es la empresa petrolera más endeudada del mundo. Ya en 2005 una gran proporción del presupuesto de Pemex, se dedicó al pago los proyectos de impacto diferido en el Registro de Gasto o Pidiregas: tres mil millones de un total de 10 mil millones de dólares (Saxe-Fernández, 2008).

A últimas fechas, Pemex ha realizado alianzas con empresas extranjeras en busca de recursos a través de los contratos de servicios múltiples, así como alianzas estratégicas para la exploración, perforación y extracción de petróleo en aguas profundas, por ejemplo, "Pemex adjudicó contratos multianuales para la explotación del gas no asociado de la Cuenca de Burgos localizada entre Tamaulipas, Coahuila y Nuevo León, que fue dividida en siete bloques: Repsol-YPF[14] se encarga del bloque Reynosa-Monterrey; a Teikoku Oil de Japón y Grupo Diavaz de México se les asignaron Cuervito y Fronterizo. Tecpetrol, domiciliada en Argentina, junto a Industrial Perforadora de Campeche, ganaron el contrato para la Misión. Grupo Lewis Energy de Texas ganó Olmos. Se asignó Pandera-Anáhuac a Industrial Perforadora de Campeche y Compañía de Desarrollo de Servicios Petroleros. En 2005, se entregó Pirineo y Monclova" (Saxe-Fernández, 2008: 7).

Empresas como Repsol se encargan de los grandes negocios del gas (regasificación, distribución, etc.), en la frontera norte. En Altamira están Royal Dutch Shell, Total y Mitsui con la participación de Unión FENOSA de España, en la Costa Azul, Sempra Energy y Royal Dutch Shell; en la Península de Baja California, cerca de las Islas de Coronado, Chevron-Texaco; en Lárazo Cárdenas, Repsol-YPF; en Tijuana, Maratón con Golar LNG Limited y Grupo GGs; en Sonora, DKRW Energy (Saxe-Fermández, 2008).

Para Rodríguez (2005), Pemex carece de marco legal para hacer alianzas con otras petroleras, cualquier tipo de exploración y producción implica compartir los resultados de la producción o en valor, situación que no permite la Constitución en su artículo 27: "las alianzas tecnológicas son una ficción y están diseñadas para compartir los resultados de la exploración y explotación" (Rodríguez, 2005: 5). Para poder realizar estas asociaciones se tendría que hacer modificaciones a la Constitución y en caso de que sucediera se allanaría el camino para entregar concesiones.

De acuerdo con David Ibarra, el proceso de privatización ya está en curso: "La privatización de las operaciones de compra y venta y transporte de gas, la venta de las instalaciones de petroquímicas, los contratos de servicios múltiples, el desplazamiento del Instituto Mexicano del Petróleo por servicios externos de asesoría, y el *outsourcing* de otras funciones –alquiler de barcos, plataformas e instalaciones, por ejemplo– son otros tantos casos de la fragmentación deliberada de Pemex y de las transferencias de oportunidades de negocios principalmente al sector privado del exterior" (Ibarra, 2008: 19).

La privatización periférica de Pemex sigue manteniendo la producción bajo propiedad estatal, sin embargo, las alianzas estratégicas con empresas extranjeras están allanando el camino a la privatización total de la empresa mexicana.

## RESULTADOS DE LA PRIVATIZACIÓN

Contrariamente a las expectativas de la política ortodoxa, la liberalización del mercado no ha mejorado los resultados económicos de México, tampoco ha creado mejores empleos ni ha mejorado los niveles de ingreso.

De acuerdo con José Luis Calva (2007), durante el periodo 1982-1987 se redujo la inversión pública y el gasto público programable, que trajo consigo el achicamiento del Estado a través de la privatización o liquidación de empresas públicas, que la inversión pública se redujo de 10.4% del PIB en 1982 a 4.9% en 1988; el gasto público para el fomento económico sectorial disminuyó 11.9% del PIB en 1982 a 8.7% en 1988.

Al concluir el sexenio, el poder de compra de los salarios se redujo en 46.8 %, la inflación alcanzó a finales de 1987, 141.8 %, el desempleo alcanzó 15 % de la población económicamente activa (PEA). El gasto social en materia de salud, disminuyó 18 % en 1985 con respecto de 1981; y en educación decreció 26 %en 1985 con respecto de 1981. "*Suma sumarum*: un sexenio de crecimiento cero (0.2% anual), que implicó una caída del PIB *per cápita* a una tasa de 2% anual" (Calva, 2007: 25). De esta manera, la década de los ochenta fue llamada la década pérdida.

En la década de los noventa la inversión pública disminuyó de 5.6% del PIB en 1987 a 3.5% en 1994; el gasto público en fomento económico sectorial pasó de 9.2% del PIB en 1987 a 4.8% del PIB en 1994. Con la crisis de 1994 el desempleo abierto alcanzó 7.6 %, con el consecuente deterioro social. Para 1994-2000, el gasto público y la inversión pública continuaron disminuyendo. Así, el gasto público cayó de 4.8 en 1994 a 3.5% en 2000, la inversión pública se redujo de 3.5% del PIB a 2.6% del PIB en el 2000 (Calva, 2007).

Durante el gobierno de Fox los resultados no fueron los esperados, por ejemplo, "durante el quinquenio 2001-2005, se logró una inflación relativamente baja de 4.9% anual; pero casi un nulo crecimiento económico del 1.8% anual durante 2001-2005, de modo que el PIB por persona apenas creció 0.7% anual, y con un persistente desequilibrio en la balanza comercial (déficit de 136 729.1 millones de dólares durante el quinquenio), así como de cuenta corriente cuyo déficit alcanzó los 56 633.4 mdd durante el quinquenio, no obstante las enormes entradas de divisas por exportación de mano de obra y los sobreprecios del petróleo" (Calva, 2007: 28-29).

Para Villegas (2010), la apertura de la economía aumentó el comercio y las inversiones extranjeras, pero la demanda de exportaciones no estimuló el crecimiento, que las actividades no vinculadas a la maquila están concentradas en sectores que no tienen muchos vínculos de demanda con el resto de la economía, como es el caso de las industrias de componentes y accesorios de autos.

En términos de desempleo, según el reporte gubernamental del segundo trimestre de 2012, la tasa de desempleo es de 4.8 %. Así de una población económicamente activa (PEA) del orden de los 50 millones de personas, la tasa de desocupación es equivalente a dos y medio millones de personas sin empleo.

Según Rojas (2012), de un total cercano a 30 millones de ocupados, correspondientes a 60 por ciento de los ocupados percibe menos de tres salarios mínimos. Cerca de 7 millones de ocupados perciben un salario mínimo o menos de uno de remuneración. Poco más de 11 millones perciben entre uno y dos salarios mínimos. Y casi 11 millones perciben entre dos y tres salarios mínimos.

El resultado de la economía mexicana demuestra que tan pronto como el gobierno permite la participación sin ninguna regulación del sector privado en la economía, el escenario será de bajas tasas de crecimiento y la pobreza y las privaciones del mercado de trabajo, seguirán siendo parte de la vida diaria. El decreto presidencial de liquidación en 2009 de la Compañía de Luz y Fuerza del Centro, que dejó a más de 40 mil trabajadores fue un paso más en la ofensiva general en contra de los movimientos sociales, en vista de que ese sindicato había sido un fuerte partidario de las protestas sociales contra las políticas neoliberales.

Ante la falta de oportunidades que genera la crisis del capitalismo han surgido varios movimientos de indignados que han ocupado las plazas principales en sus respectivos países, en distintas partes del mundo, como el movimiento 15 de mayo (15M) de 2011, en España, por las elevadas tasas de desempleo. Las revueltas en la primera árabe, los movimientos estudiantiles, principalmente en Chile y Colombia, la ocupación de la plaza de *Wall Street*, Plaza Libertad. De lugares distantes como Roma, Londres o la Ciudad de México con #yo soy 132, en lucha por la democracia, El Movimiento por la Paz con Justicia y Dignidad, en contra de la guerra con el narcotráfico que ha dejado más de 95 mil muertos, según cifras del INEGI en 2012.

En México y en todas partes del mundo donde se han indignado, los gritos comunes son: el desempleo, la falsa democracia, la precariedad impuesta, educación y la exclusión. La enorme brecha entre ricos y pobres es puesta de nuevo en la mesa. "Somos 99 %" han gritado los excluidos del planeta, mientras el acaudalado uno por ciento se aferra al mercado y al poder (Véase suplemento especial, *La Jornada*, 2011), entre los que se encuentran Bill Gates, Warren Buffet y Carlos Slim.

## DE LA PRIVATIZACIÓN A LAS FUSIONES Y ADQUISICIONES

Entre 1984 y 1987 las principales empresas mexicanas superaron su crisis de la deuda externa, debido a sus estrategias de inversión financiera, que les permitió en 1987 reemprender la inversión productiva, relegada a principios de la década. Para 1989 y 1990, a México se le abrieron las puertas de los mercados financieros internacionales, mismos que estuvieron vedados desde inicios de los ochenta, lo cual dio un nuevo impulso a la expansión de las grandes empresas y grupos (De los Ríos, 2005).

De acuerdo con Garrido (2001) en México las repercusiones y los resultados de las fusiones y adquisiciones pueden caracterizarse en macro, meso y microeconómico. En el aspecto macroeconómico resalta la exposición de la economía nacional a la competencia internacional, las reformas económicas y las privatizaciones que afectaron importantes sectores económicos y la competencia. En el mesoeconómico se observa la apertura de los sectores manufactureros y de servicios a la inversión extranjera directa. En el ámbito microeconómico, la estructura del liderazgo de las grandes empresas registró importantes cambios en el tipo de propiedad institucional, lo que incrementó la importancia de las transnacionales de las grandes empresas. Asimismo, el desempeño de las empresas, luego de los procesos de fusión y adquisiciones transfronterizas que llevaron a cabo las empresas mexicanas durante los años noventa, muestran que los casos exitosos y otros de éxito razonable, pero también de fracaso, con grandes costos para las empresas involucradas.

De 1988 a 2005, el valor de adquisiciones en México ascendió a 40.3 millones de dólares, las cuales tienen un ciclo errático, pues en 1995 se reducen las operaciones, luego repuntar en 1997, caer en 1998, incrementarse en 1999 y comenzar a atenuarse en 2002 (De los Ríos, 2005).

Las adquisiciones de las empresas en México durante los últimos veinte años se acentuaron en tres sectores: a) sector bancario y otros servicios financieros, b) telecomunicaciones, y c) alimentos y bebidas. En estos sectores se realizaron 64.7% de las adquisiciones (De los Ríos, 2005). Otras áreas han mostrado un número importante de adquisiciones de empresas como el comercio, transportes, puertos, almacenes, medios de comunicación, hotelería y restaurantes (Garrido, 2001).

De acuerdo con Víctor Livio de los Ríos (2005), el volumen de adquisiciones en el periodo de 1986-2008 se explican por varias razones entre ellas:

- Durante la década de los ochenta y los noventa se realizó en México el mayor número de desincorporaciones de empresas del sector público. El proceso privatizador que se inició a mediados de los ochenta tuvo su mayor actividad en el gobierno de Carlos Salinas de Gortari de 1989-1993, pues en ese lapso se vendieron las principales empresas industriales y los bancos, fundamentalmente 1990 y 1992.
- En años posteriores, el incremento de las adquisiciones fue consecuencia indirecta del proceso de privatización, ya que muchas empresas públicas en primera instancia fueron adquiridas por empresas e instituciones mexicanas, posteriormente pasaron a manos del capital transnacional, como el sector bancario.
- El proceso de apertura y desregulación en importantes sectores económicos como el bancario-financiero y el de telecomunicaciones, implicaron a finales de los años noventa y principios del siglo xxi, un importante número de mega-fusiones y adquisiciones.

Las adquisiciones de empresas en México durante los últimos veinticinco años se observa que las operaciones más importantes se realizaron en tres sectores: 1) sector bancario y otros servicios financieros, 2) telecomunicaciones, y 3) alimentos y bebidas.

Durante el gobierno de Carlos Salinas de Gortari en el sistema financiero y en particular la banca, se propició la integración de grupos o consorcios en manos del capital del país. Por ejemplo, las privatizaciones de empresas públicas realizadas durante el gobierno de Salinas de Gortari consideraban las modificaciones legales y el proceso mismo de venta de los activos, como medios para permitir que las nuevas firmas fueran adquiridas y controladas por el país (Vidal, 2001). Para hacer posible la medida se emitió una nueva Ley de Instituciones de Crédito que además de establecer la transformación de los bancos en sociedades anónimas, aseguraba que la banca del país fuera controlada por capitales mexicanos.

Sin embargo, con la crisis bancaria y monetaria de 1997 y los serios problemas de cartera vencida de una gran cantidad de bancos se producen nuevas modificaciones legales que terminan por eliminar las restricciones a la participación del capital extranjero en la banca y todo el sistema financiero. En 1998, el Congreso aprobó la Ley de Protección al Ahorro Bancario, con ella se dio la apertura total en banca y bolsa al capital extranjero, eliminándose las restricciones en el caso de los tres mayores grupos bancarios del país (Vidal, 2008).

Por su parte, en Tratado de libre Comercio de América del Norte (TLCAN) hay nuevas limitaciones o excepciones. Destacan las reglas específicas para considerar como producción en la zona del TLCAN, a la industria automotriz, eléctrica y electrónica. Los cambios realizados en diversas leyes, la aplicación flexible de otras, el establecimiento de reglamentos y los propios cambios en la gestión de la economía por parte del Estado, han dado como resultado la eliminación de las restricciones a la acción del capital extranjero en el país (Vidal, 2008).

En la Banca y el sistema financiero, también se destaca el peso de las empresas extranjeras. Los cambios en las leyes, reprivatizaciones, cambios en los principales accionistas, continuas adquisiciones y ventas de capital extranjero, dan como resultado que en la banca cinco de los seis mayores están en manos de la banca transnacional y en 2005 ya tenían 80% de los activos. Así tenemos, a BBVA-Bancomer, Banamex-Citigroup, Santander, HSBC y Scotiabank-Inverlat. Todos ellos son grupos financieros que lo mismo actúan en los mercados de dinero y capital, que financiando vivienda residencial.

En algunos casos, los empresarios de México han tenido la opción de canjear sus acciones por participaciones en el capital de la firma extranjera adquiriente. Por ejemplo, en la compra de Banamex-Accival por Citigroup, por un monto de 12 mil 500 millones de dólares. De esta manera, los empresarios mexicanos quedan en condiciones de socios muy minoritarios de la matriz que adquiere la empresa en México, haciendo suyos los intereses de la firma. Actualmente 90 % del sistema financiero mexicano está en manos de instituciones extranjeras; Banorte[15] es el único banco mexicano bajo control de sus propietarios originales (Flores, 2002).

En este contexto, los usuarios enfrentan una banca extranjera que aprovecha sus nichos de rentabilidad cobrando extraordinarios comisiones, por los servicios que les prestan y no cargan los niveles de precios por los mismos servicios en sus países de origen.

Para Bouzas Ortiz (1998), las alianzas estratégicas llevan a renunciar a la banca nacional a sus propios proyectos de modernización y adoptar las directrices impuestas por los socios inversionistas extranjeros. De tal suerte, que en la década de los noventa, no sólo se reprivatiza la banca, sino que con todas las reformas a leyes necesarias, incluso constitucionales, se entrega la banca a los bancos extranjeros incorporando a los banqueros nacionales a la globalización financiera como socios sometidos, como representantes criollos de un proyecto no nacional.

En síntesis, la oligopolización extranjera del sistema bancario mexicano tiene menos compromiso con el desarrollo del país que la propia banca nacional; por lo que no es recomendable que la participación del capital foráneo en el sistema bancario sea mayor que el nacional, sin embargo, en la actualidad la extranjerización bancaria está consolidada.

Por otra parte, la asociación con el capital extranjero se presenta de otras formas. Por ejemplo, en Gruma participa la empresa estadounidense Archer-Daniels-Midland. Sus relaciones son múltiples, incluso en Molinera México comparten capital en una proporción de 60% para Gruma y 40% para Archer-Daniels-Midland. En grupo modelo participa la firma Anheuser-Busch (AB) que está entre las dos primeras empresas cerveceras del mundo, con compañías en diversos países, una marca que comercializa a nivel global y asociaciones en Europa como en Asia. Mientras que Modelo tiene sólo empresas en México y la mayor parte de sus exportaciones son en Estados Unidos, mercado en el que AB es la empresa líder (Vidal, 2008).

De acuerdo al informe de la consultora especializada en fusiones y adquisiciones de empresas Pablo Rión y Asociados, señala que en 2008 las operaciones sumaron 18 mil 532 millones de dólares, 48% menos que los 38 mil 517 millones del año anterior. (Véase cuadro 4).

En el reporte se destaca que las empresas mexicanas fueron las más dinámicas en el mercado, al protagonizar dos terceras partes del número de operaciones; es decir, 112 de los 182 casos en que compañías nacionales aparecen como adquirientes, ya sea en el país o en el extranjero. Véase cuadro 5.

Comparado con el año anterior, en 2008 se contrajo el número de operaciones 5%, al pasar de 192 a 182 en 2007, la cifra es casi equivalente al resultado de 2006, cuando se realizaron 184 operaciones, y se compara favorablemente con las 121 del 2005 y 107 del 2004.

De las principales transacciones de compañías públicas en México, la venta de Cimpor Inversiones por Cemex, la venta hecha por Telmex del operador de cable en Colombia Superview, la adquisición de 28% de Circuit City Soteres realizada por Ricardo Salinas Pliego, la compra de Consorcio Grupo Hotelero T2 hecha por el Grupo Aeroportuario Centro Norte. Asimismo Comercial Mexicana vendió 50% de su participación en Prestacomer a su socio el brasileño BNP Paribas Finance; Grupo Collado adquirió la participación que no poseía en Coryer a su hasta entonces socio Ryerson; Prolec-GE, subsidiaria de Xinux, compró 54.4% de la firma india Indo Tech Transformer, y la familia Aramburuzabala (que forma parte de los accionistas principales del cervecero Grupo Modelo), adquirió una participación adicional en Ingenieros Civiles Asociados (ICA) para alcanzar 5.1% del capital accionario.

*Cuadro 4*
Resumen de transacciones por monto de sector

| Sector/Industria | Monto de las transacciones (mdd) | | | | |
|---|---|---|---|---|---|
| | *2004* | *2005* | *2006* | *2007* | *2008* |
| Construcción | $ 8,077 | $ 1,311 | $ 1,630 | $19,469 | $ 2,242 |
| Consumo | $ 1,553 | $ 2,219 | $17,229 | $ 9,712 | $ 8,528 |
| Industria /manufactura e ingeniería | $ 4,688 | $ 6,407 | $ 3,297 | $ 4,381 | $ 1,232 |
| Medios | $ - | $ - | $ - | $ 1,627 | $ 4,890 |
| Salud, ciencias y medicina | $ 4 | $ 106 | $ 294 | $ 1,254 | $ 245 |
| Servicios financieros | $ 165 | $ 69 | $ - | $ 1,240 | $ 257 |
| Servicios profesionales | $ 4,335 | $ 290 | $ 1,667 | $ 423 | $ 103 |
| Tecnología | $ 17 | $ 57 | $ 29 | $ 378 | $ 662 |
| Telecomunicaciones | $ 1,581 | $ 3,035 | $ 6,437 | $ 24 | $ 363 |
| Transporte | $ 705 | $ 4,854 | $ 2,009 | $ 10 | $ 5 |
| Turismo y entretenimiento | $ - | $ 88 | $ 132 | $ - | $ 10 |
| Total | $ 21,125 | $18,436 | $32,723 | $38,517 | $18,532 |
| Transacciones atípicas por tamaño (1) | $ - | $ - | $ 14,000 | $15,300 | $ - |
| Monto normalizado | $ 21,125 | $18,436 | $18,723 | $23,217 | $18,532 |

(1) Para normalizar los montos de 2006 y 2007, se separó la fusión de Pernod Ricard y Allied Domecq (2006) y la adquisición de Rinker por Cemex (2007).
Fuente: Rión Pablo y Asociados (2011: 1).

*Cuadro 5*
Resumen del número de transacciones por nacionalidad del adquiriente

| Nacionalidad | 2006 | 2007 | 2008 |
|---|---|---|---|
| | *Número* | *Número* | *Número* |
| Mexicana-Mexicana | 73 | 75 | 56 |
| Mexicana-Extranjera | 48 | 34 | 56 |
| Extranjera-Mexicana | 51 | 62 | 50 |
| Extranjera-Extranjera | 12 | 21 | 20 |
| Total | 184 | 192 | 182 |

Fuente: Rión, Pablo y Asociados (2011: 1).

De las operaciones privadas destacaron la compra de acciones de JP Morgan Grupo Financiero en México por parte de The Bank of New York Mellon; el principal accionista de la Universidad ICEL, pactó la compra del periódico *El Economista* por 25 millones de dólares. La estadounidense Delphi vendió su negocio de escapes a la empresa mexicana Bienes Turgón; AMC Entertainment vendió Grupo Cinemex a Entretenimiento GM de México, en 237.5 millones de dólares. La francesa Vallee Group adquirió la unidad de teléfonos públicos de Ascom con operaciones en Francia y México; la austriaca Salzburger Aluminium compró una participación de la mexicana Mecasa, productora de tanques de combustible para vehículos pesados.

La holandesa Royal Boskalis Westminster compró a su socio mexicano el 50 % de sus acciones de Dragamex, una empresa especializada en dragado. La telefónica regiomontana Marcatel adquirió a la canadiense Aldea Visión. La japonesa Funai Electric compró a Phillips North America el negocio de Blu-ray, DVD, discos duros y teatros, operación que incluyó a las plantas ubicadas en el norte del país. Grupo Financiero Monex. Komet of America adquirió de la suiza Klingelnberg Oerlikon, la empresa Afilatec, ubicada en Querétaro. El empresario Carlos Slim Helú incrementó su participación accionaria en la cadena estadounidense Saks. MVS Multivisión formó una empresa conjunta con la estadounidense EchoStar Corporation para ofrecer televisión satelital con el nombre de Dish México (Rión, 2011).

A pesar de la crisis de las hipotecas sub-prime en 2009, las fusiones y adquisiciones continuaron. La cervecera holandesa Heineken adquirió intereses cerveceros de la firma Femsa por un total de 7,374 millones de dólares.[16]

En diciembre de 2009, Walmart de México (Walmex) adquirió el 100 por ciento de Walmart Centroamérica. Walmart Stores ya poseía el 51 por ciento y compró el 49 por ciento restante de los accionistas locales. La transacción de este 49 por ciento se estima en 1,345 millones de dólares combinando el pago de efectivo y acciones de Walmex.

En 2010, se dio la compra de las acciones del Grupo Carso Telecom por 27,400 millones de dólares (20,000 millones de euros) por parte de América Móvil, empresa mexicana de telecomunicaciones, de la que es propietario Carlos Slim. La firma se hizo con 99,4% de los activos del Grupo Carso, controlador de Telmex y Telmex Internacional.

Como se puede observar, las empresas mexicanas fueron muy dinámicas en el mercado en la primera década del siglo XXI, controlando sectores importantes de la economía como la construcción, telecomunicaciones, turismo, etc., sin embargo, con el tiempo algunas empresas pasaron a manos del capital transnacional, como el duopolio cervecero Grupo Modelo y Femsa, adquirido por el consorcio belga *Anheuser-Busch InBev* y la firma holandesa *Heineken*, respectivamente.

**Conclusiones**

El análisis de la evolución del proceso de privatización, fusiones y adquisiciones de las grandes empresas en México, en las últimas décadas, indican que éstas han favorecido a las empresas transnacionales, debido entre otros cambios a los realizados en diversas leyes, su aplicación flexible, la retracción del Estado en la economía, eliminando las restricciones a la acción del capital extranjero en el país.

En principio, muchas empresas públicas fueron adquiridas por firmas e instituciones mexicanas, posteriormente pasaron a manos del capital transnacional, teniendo un efecto indirecto en el incremento de las fusiones y adquisiciones, como el sector bancario, que está en manos de la banca extranjera que controla cinco de los seis mayores en el país bbva-Bancomer, Banamex-Citigroup, Santander, hsbc y Scotiabank-Inverlat.

Contrariamente a las expectativas de la política ortodoxa, la liberalización del mercado no ha mejorado los resultados económicos de México, tampoco ha creado mejores empleos ni ha mejorado los niveles de ingreso.

En términos de desempleo, según el reporte gubernamental del segundo trimestre de 2012 la tasa de desempleo es de 4.8%. Así de una población económicamente activa (pea) del orden de los 50 millones de personas, la tasa de desocupación es equivalente a dos y medio millones de personas sin empleo, a ello se suma el índice de pobreza que hemos tenido en estos años de más de 50 millones de personas que viven en la zozobra y ven el futuro con incertidumbre. Sólo unas cuantas familias mexicanas y capitales extranjeros se han beneficiado.

Con esto no quiero decir, que regresemos al Estado omnipresente, sino que allí donde la empresa y el gobierno han demostrado un comportamiento deficitario irreparable, la privatización junto con la liberalización y la desregulación pueden intervenir para traer beneficios, pero no considero que el Estado deba dejar de intervenir donde lo reclama el bienestar social. En la actualidad hay actividades que no son ajenas al Estado como la infraestructura, la educación, la salud, la seguridad, la justicia, etcétera. Deben fortalecerse éstas a través de la inversión pública para erradicar la pobreza, no se debe renunciar a ellas por capricho del mercado.

**Notas**

[1] Con el vocablo global aludo al proceso de integración planetaria que viene desarrollando el capital, en tanto relación de producción, es decir, como el proceso de valorización del capital y de la necesidad, por tanto, del proceso de reproducción ampliada.

[2] El Estado-nación es literalmente una soberanía dirigida por sola una nación. De algún modo es una creación ideológica que alude a la formación de un poder político bien constituido, con

identidad, cultura, lenguaje, costumbres, código moral, formas de recreo, ideas éticas, etcétera. Como bien dice Ayala Espino (1997), en rigor no existe un Estado-nación ideal y probablemente nunca haya existido, pero considera que ello no quiere decir que los gobiernos no se propongan desarrollar y/o fortalecer un Estado-nación, tarea cada vez más difícil en la era de la globalización en la cual la misma noción de soberanía ha quedado cuestionada.

3   "Las políticas de liberalización financiera tienen su origen en los postulados de la teoría neoclásica y en los planteamientos desarrollados por Ronald Mckinnon y Gurley Shaw. Ambos economistas proponen la necesidad de concentrar la atención en el desarrollo de los mercados de capital por lo que es necesario la eliminación de todo tipo de *represión financiera* ya que tenía efectos negativos en la generación del ahorro, su movilización y asignación eficiente, siendo necesario eliminar todo tipo de interferencia por parte del gobierno o Estado y otorgando esta responsabilidad al mecanismo de mercado" (Lucero, 1999: 92).

4   "Si la globalización financiera propone el acercamiento de los distintos sistemas financieros operantes implica la generación de servicios financieros homogéneos, un conocimiento perfecto de los distintos mercados, la libre movilidad y capacidad de los intermediarios para dirigirse y ubicarse en cualquier parte del mundo donde la posibilidad de asociaciones, coinversiones o fusiones son los principales elementos para adquirir presencia inmediata en cualquier parte del mundo. La unificación de los servicios financieros en el plano mundial es igualmente conocida con diferentes nombres: *internacionalización de los servicios financieros, integración de los mercados financieros a escala mundial o internacional* y, por último, *universalización de los servicios financieros*" (Lucero, 1999: 92)**.**

5   Las economías emergentes se clasifican en esta forma debido a que sus mercados financieros no están del todo desarrollados. Esto último se refiere a la capacidad que tiene el mercado para que la información llegue a todos los inversionistas y que esto sea registrado en los precios.

6   Cursiva mía.

7   "El sector público es el ente encargado de realizar las actividades políticas y administrativas –programables y medibles en su connotación macroeconómica, administrativa, contable y social– requeridas para alcanzar el fin fundamental del Estado: constituirse en un instrumento efectivo de promoción humana, social e individual, tanto en el plano político-social como en el económico" (Ayala, 1997: 44). Para Inostroza Fernández (1990), el sector público abarca una serie de acciones que se realizan en tres planos. En el plano del gobierno y la administración pública determinado país. En el plano de la actividad productiva de las empresas públicas, en las áreas estratégicas y prioritarias del desarrollo nacional. Y en el plano institucional del sector público determinado por la administración pública, que la integran el gobierno central, el sector paraestatal (empresas públicas) y los bancos o entidades financieras del Estado.

8   Este tipo de compra ocurre cuando un pequeño grupo de accionistas, utilizando dinero prestado, obtenido frecuentemente con bonos de alto riesgo, se apoderan de una compañía.

9   Los organismos descentralizados son las personas jurídicas creadas o prioritarias, para la realización de actividades estratégicas y prioritarias, para la prestación de un servicio público o social y para la obtención o aplicación de recursos para fines de asistencia o seguridad social.

10   Las empresas de participación estatal mayoritaria, son las que determinan como tales la Ley Orgánica de la Administración Pública Federal (LOAPF), y en las que el gobierno o una o más entidades paratestales participen con más de la mitad del capital nacional (Art. 29).

11   Los fideicomisos públicos que se establezcan, se organizan de manera análoga a los organismos descentralizados o empresas de participación estatal mayoritaria, que tengan como propósito auxiliar al ejecutivo mediante la realización actividades prioritarias (Art. 40).

12   Para una mejor comprensión del proceso de privatización de Telmex, en lo particular, véanse Telmex (1992), *La desincorporación de Teléfonos de México*. Aspe Armella, Pedro (1993), *El cambio mexicano de la transformación económica*. México, FCE. Rogozinski, Jacques (1993), *La privatización de empresas paraestatales*, México, FCE. Sacristán Roy, Emilio (2006), "Las privatizaciones en México", en *Economía*, Vol. 3, No. 9, México, UNAM.

13   La mayoría de las operaciones de venta que se han formalizado en esta administración han quedado en manos de mexicanos, para ser precisos 93%.

14   Los hidrocarburos se introdujeron en las exportaciones mexicanas a España en 1971, con la exportación de petróleo crudo a la Compañía Española de Petróleos, S.A. (Cepsa). En 1979, con la entrada en vigor del Acuerdo de Cooperación Industrial, Energética y Minera, Pemex adquirió una participación accionaria en Petróleos del Norte, S.A. (Petronor). En 1990, en el marco de un acuerdo de Cooperación estratégica, se procedió a un intercambio de acciones de Petronor por acciones de Repsol-YPF, S.A., mediante el que Pemex adquirió 2.88% del capital social de la petrolera española. En 1992, Pemex alcanzó un total del 5% de capital social de Repsol-YPF, S.A. Actualmente Repsol es uno de los socios más importantes de Pemex (Carmona y Pérez-Noyola, 2011).

15   En 2010, el Grupo Financiero Banorte, alcanzó un acuerdo definitivo con los accionistas de Ixe Grupo Financiero para integrar ambas organizaciones a través de compra del 100 por ciento del paquete accionario de esta última a cambio de la emisión de acciones representativas del 13 por ciento del nuevo capital de Banorte. La operación, en función del valor de la acción al momento del anuncio, valorizó a Ixe en 1,330.9 millones de dólares.

[16] Por su parte, el Grupo Modelo fue adquirido en junio de 2012 por el consorcio belga *Anheuser-Busch InBev*, por la suma de 20 mil 100 millones de dólares. Con ello, se dio fin al duopolio cervecero mexicano (Femsa se vendió a Heineken) dando paso al duopolio cervecero transnacional. En esta compra dejaron de pagar impuestos 6 mil millones de dólares (Fernández-Vega, 2012).

**BIBLIOHEMEROGRAFÍA**